# High-Speed Femto-Joule per Bit Silicon-Conductive Oxide Nanocavity Modulator

Erwen Li, Bokun Zhou, Yunfei Bo, and Alan X. Wang[1], *Senior Member, IEEE*

*Abstract*— By combining the large Purcell effect of photonic crystal nanocavity and the strong plasma dispersion effect of the transparent conductive oxides, ultra-compact silicon modulators with heterogeneously integrated indium-tin-oxide (ITO) can potentially achieve unprecedented energy efficiency to atto-joule per bit. In this article, we report the first high-speed silicon nanocavity modulator driven by an ITO gate, achieving 2.2 GHz bandwidth. On-off-key modulation is measured up to 5 Gb/s with only 2 V voltage swing and 18.3 fJ/bit energy efficiency. In addition, we perform in-depth analysis of the energy efficiency and high frequency simulation of the nanocavity modulator, revealing the critical role played by the semiconductor conduction path and the overlapping factor between the accumulated free carriers and the cavity resonant mode. Based on our analysis, we propose a strategy to further improve the modulation bandwidth to 23.5 GHz by node-matched doping and reduce the energy consumption to the range of hundreds of atto-joule per bit.

*Index Terms*— electro-optic modulators, optical interconnections, optical resonators, photonic crystals, silicon photonics, transparent conductive oxides

## I. INTRODUCTION

Silicon photonics provides exclusive advantages for on-chip optical interconnects in terms of bandwidth density, latency, energy efficiency, and cost effectiveness [1, 2]. Micro-resonators play pivotal rules in silicon photonic devices including high-speed and low-energy-consumption electro-optical (EO) modulators [3, 4], wavelength-division multiplex/demultiplex (WDM/DeMUX) modules [5], and chip-scale optical input/output (I/O) systems [6]. The light-matter interaction can be dramatically enhanced by utilizing a resonator structure such as micro-ring [3, 7], micro-disk [4], or photonic crystal (PC) nanocavity [8-10]. Compared with conventional Mach-Zehnder interferometer (MZI) modulators [11], resonator-based modulators offer remarkable reduction in device size, driving voltage and energy consumption, which can meet the stringent requirement of high density on-chip optical interconnects. Especially, as the bandwidth density of optical interconnect increases exponentially, the energy consumption per bit becomes the primary concern of the whole system. High energy efficiency of photonic devices can be achieved by increasing the Purcell factor of resonator modulators [12]. Among existing micro-resonators, PC nanocavity can offer much larger Purcell factor compared with micro-ring or micro-disk resonators [13] due to the ultra-compact mode volume. High Purcell factor resonator modulator is the only feasible design to simultaneously achieve high energy efficiency and large bandwidth as proved in previous studies [14].

In addition to the development of resonator-based E-O modulators, heterogeneous integration with other functional materials shows great promise to overcome the intrinsically weak plasma dispersion effect of silicon. Various active materials such as III-V semiconductors [15, 16], 2D materials [17-19], transparent conductive oxides (TCOs) [9, 10, 20-24], $LiNbO_3$ [25, 26], and phase change materials [27, 28] have been integrated on silicon to build high performance photonic devices. In recent years, indium-tin oxide (ITO) as a type of transparent conductive oxide (TCO) materials are attracting escalating attention for EO photonic devices due to their strong plasma dispersion [29]. Moreover, TCOs exhibit unique epsilon-near-zero (ENZ) effect in the telecom wavelength range [20-23], which can be used to enhance the free carrier absorption. Ultra-compact and broadband electro-absorption (EA) modulators using ENZ effect based on plasmonic slot waveguide [22, 23] and hybrid plasmonic-silicon waveguide structure [20, 21] have been demonstrated in recent years. Besides, moderately doped TCOs are electrically conductive but optically transparent, which makes them perfect gate materials for heterogeneous integration with silicon micro-resonators. Extremely large wavelength tunability of 250 pm/V has been demonstrated on both TCO-gated PC nanocavity [9], [10] and micro-ring resonator [24], which shows great potential for low-voltage and ultra-energy-efficient optical modulators and filters. Therefore, combining the large Purcell factor of PC nanocavity and the extremely large EO efficiency of TCO gates can potentially enable nanophotonic modulators with atto-joule/bit energy efficiency and high modulation bandwidth [10]. However, the modulation speed of previous TCO resonator modulators was limited to a few Mega-Hertz range due to the large series resistance and the lack of high-speed electrode design. Only TCO-based EA modulators achieved a data rate of 2.5 Gb/s at the cost of high driving voltage and energy consumption [21].

In this paper, we conducted holistic design of ultra-compact, high-speed PC nanocavity modulator driven by an ITO gate

This paragraph of the first footnote will contain the date on which you submitted your paper for review. This work is supported by the AFOSR MURI project FA9550-17-1-0071 and the NSF GOALI project 1927271. *(Corresponding author: Alan X. Wang.)*

E. Li, B. Zhou, Y. Bo and A.X. Wang are School of Electrical Engineering and Computer Science, Oregon State University, Corvallis, OR 97331 USA (e-mail: wang@engr.orst.edu).

based on silicon rib waveguide and high-speed coplanar electrodes. In-depth analysis of the energy efficiency and high frequency simulation of the nanocavity modulator reveals the critical role played by the semiconductor conduction path and the overlapping factor between the accumulated free carriers and the cavity resonant mode. To the best of our knowledge, this is the first systematic analysis of any high-speed TCO photonic devices. Experimentally, we achieved a 3-dB bandwidth of 2.2 GHz. Digital modulation using on-off-key (OOK) was measured up to 5 Gb/s with only 2 V voltage swing, which yielded an ultra-high energy efficiency of 18.3 fJ/bit. Moreover, we proposed a new method using node-matched doping and high-mobility TCO gate material, which can extend the modulation bandwidth to 23.5 GHz. Based on the quantitative analysis of the overlapping factor, we proposed a strategy to improve the energy efficiency to hundreds of atto-joule per bit level.

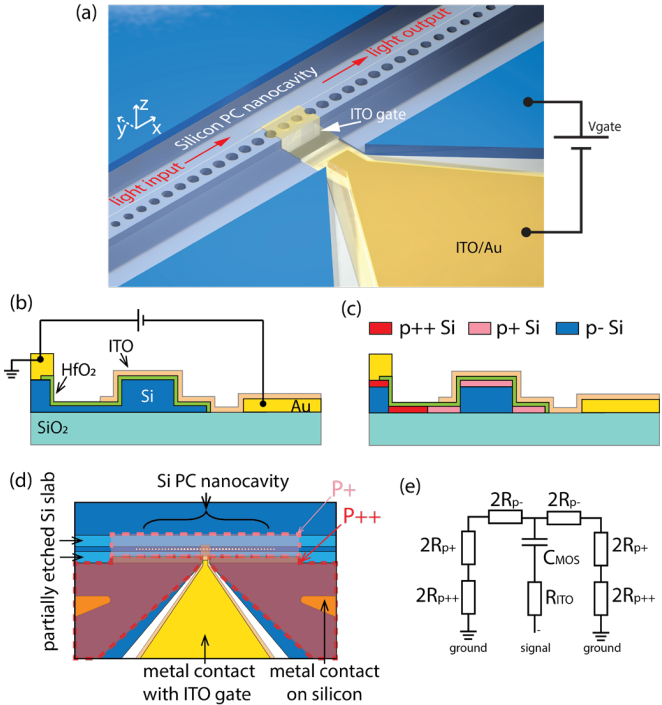

Fig. 1. (a) 3D schematic of the ITO-gated silicon PC nanocavity modulator. (b) Cross-sectional schematic of the ITO/HfO$_2$/Si MOS capacitor in the active region. The Si bottom and ITO gate contact pads are drawn on two sides of the waveguide for the ease of illustration. (c) Cross-sectional schematic of the silicon doping profile considered in the HFSS simulation. (d) Partial layout of the patterns used for the fabrication of the nanocavity modulator. (e) Equivalent circuit model of the PC nanocavity modulator in the active region.

## II. DESIGN AND FABRICATION

The working principle and fundamental device design of the PC nanocavity modulator has been described using fully etched silicon strip waveguide but can only operate at low E-O modulation speed [9-10]. Fig 1a shows the 3D schematic of the high-speed device design using silicon rib waveguide to achieve reduced series resistance in this work. The PC nanocavity consists of two back-to-back PC mirror segments on a silicon rib waveguide with 500 nm width and 250 nm height.

The active region of the nanocavity modulator consists of an ITO/HfO$_2$/Si MOS capacitor. The cross-sectional schematic is shown in fig 1b. Here the 50nm thick silicon slab waveguide serves as both the bottom of the MOS capacitor and electrical conduction path. The PC nanocavity is covered by 16 nm thick HfO$_2$ gate oxide layer. On top of that is the 20 nm thick ITO gate layer covering the center active region. When a negative bias is applied to the ITO gate, electrons and holes accumulate at both the ITO/ HfO$_2$ and Si/ HfO$_2$ interfaces, respectively. The accumulated free carriers induce blue-shift of the resonance peak, modulating the light transmission. The relationship between resonance frequency detuning and accumulated free carriers can be described by the cavity perturbation theory [14]:

$$\frac{\Delta\omega}{\omega} = -\frac{\int \Delta\varepsilon |E|^2 dv}{2\int \varepsilon |E|^2 dv} = \frac{\alpha K Q_{tot}}{2V_m}. \quad (1)$$

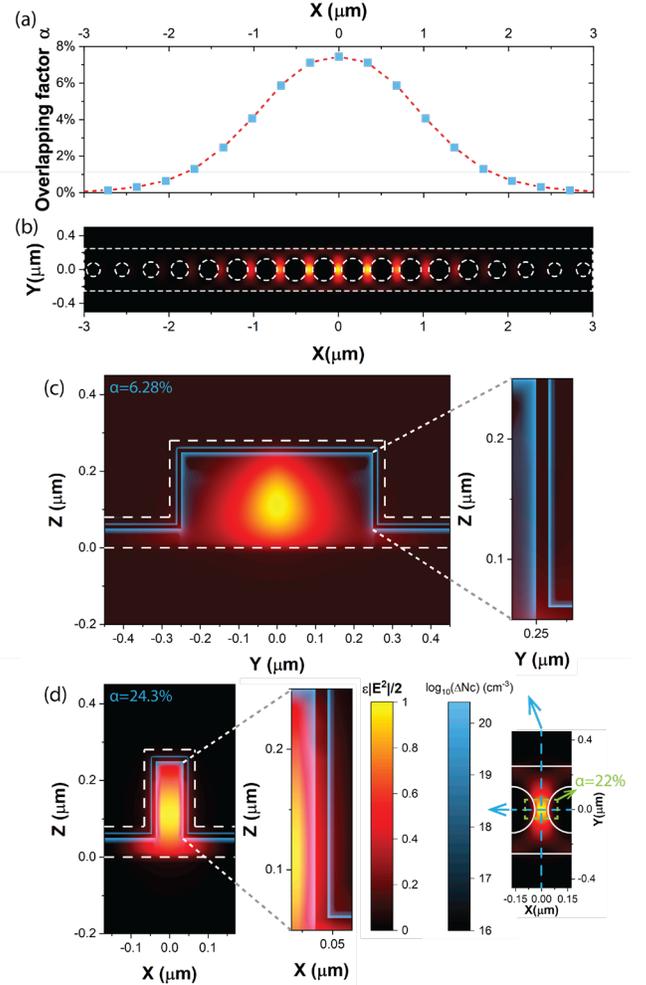

Fig. 2. (a) Overlapping factor vs. position of the ITO/HfO2/Si MOS capacitor by assuming the MOS capacitor covering 1 period of the PC nanocavity. (b) Top view of the electrical energy density distribution of the PC nanocavity. (c) and (d) Cross-sectional view of energy density distribution at (c) X=0 μm and (d) Y=0 μm with overlay of simulated free carrier density accumulation distribution at -2V (blue shadow). Inset: Zoomed-in top view of nanocavity mode profile in the center period. Blue dashed lines label the cross-section positions of (c) and (d). Green dashed box labels the region where the overlapping factor reaches 22%.

Here, E is the electric field distribution of the cavity mode, ε and Δε are the initial and perturbed permittivity distribution. Because the plasma dispersion induced permittivity change is linear to the change of free carrier density, the second term of (1) can be rewritten into the third term, where there are four major factors: $K = -\partial(\Delta\varepsilon/\varepsilon)/\partial(\Delta N_c q)$ is the free carrier dispersion coefficient, which equals to the relative permittivity perturbation per unit change of free carrier charge; $Q_{tot}$ is the total change of the accumulated carrier charge; $V_m$ is the optical mode volume of the cavity mode; and α is the overlapping factor that describes the overlapping between the free carrier perturbation and the electrical energy of the cavity mode, which can be expressed as:

$$\alpha = \frac{\int \Delta N_c q \varepsilon |E|^2 \, dv}{Q_{tot} \max(\varepsilon |E|^2)}. \quad (2)$$

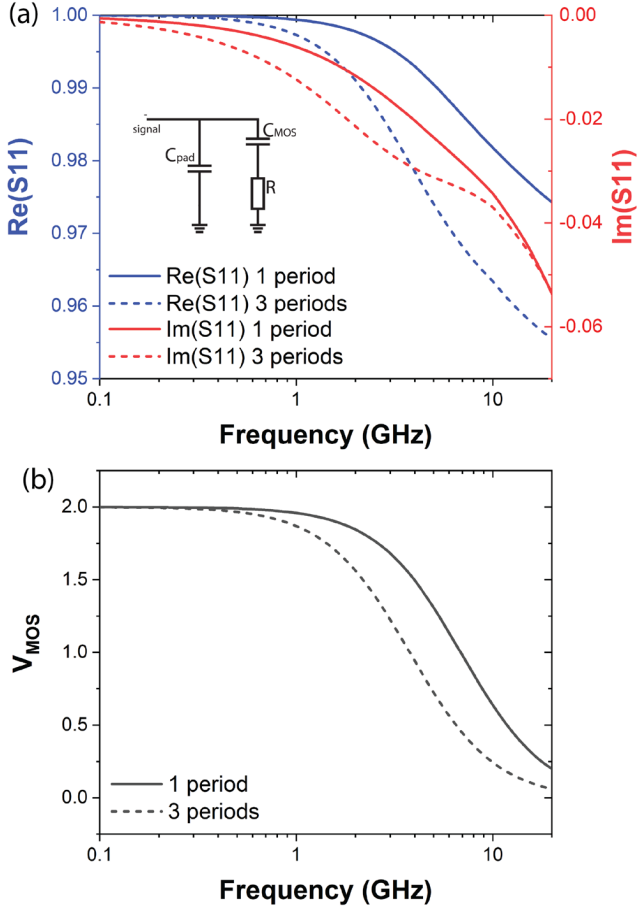

Fig. 3. (a) S11 parameters from the HFSS simulation. Inset: circuit model used to fit the S11 parameters. $C_{pad}$ is the parasitic capacitance between the contact pads. (b) Normalized voltage across the MOS capacitor as a function of frequency.

TABLE I
CIRCUITS PARAMETERS EXTRACTED FROM HFSS SIMULATION

|  | p- | p+ | p++ | ITO | $R_r$ |
|---|---|---|---|---|---|
| dopant density (cm$^{-3}$) | $1\times10^{17}$ | $5\times10^{18}$ | $1\times10^{20}$ | $2\times10^{20}$ | \ |
| Sheet resistance (Ω/□) | 10.7k (200nm) | 2.9k (50nm) | 253 (50nm) | 625 (20nm) | \ |
| Aspect ratio (3 periods) | 0.08 | 0.18 | 0.9 | 0.892 | \ |
| R (Ω) (3 periods) | 856 | 530 | 228 | 558 | 78 |
| Aspect ratio (1 period) | 0.08 | 0.22 | 0.55 | 3.19 | \ |
| R (Ω) (1 period) | 856 | 638 | 139 | 1994 | 93 |

In our previous publications [9-10,13], we have already discussed the advantages of the TCO-Si PC nanocavity modulator for achieving ultra-low energy consumption from perspectives of active material, optical resonator and capacitor design. However, in order to maximize the energy efficiency of the modulator, optimizing the overlapping factor α is also equally important. The overlapping factor determines the percentage of the electrically accumulated free carriers that can be used to interact with the light. Fig 2b plots the top view of the normalized electrical energy density ($\varepsilon|E|^2/2$) distribution of the PC nanocavity simulated by 3D finite-difference time-domain (FDTD) using Lumerical FDTD, which corresponds to an ultra-compact optical mode volume of 0.058 μm$^3$ ($0.66(\lambda/n_{Si})^3$). We can estimate the overlapping factor from the electrical energy distribution under the assumption of 1nm thick uniform accumulation layer. Fig 2a plots the calculated overlapping factor versus the position of the active TCO-Si MOS capacitor by assuming the MOS capacitor covering one period of the PC nanocavity. Clearly, covering 3 periods of the PC nanocavity can provide a balance between energy efficiency and modulation speed. Fig 2c and 2d show the cross sectional electrical energy density distributions of the center period of the nanocavity modulator at positions X=0 μm and Y=0 μm, respectively with overlays of accumulated free carrier distribution after applying -2V bias on the ITO gate, which was simulated by Silvaco. We can directly visualize how the accumulated carriers are overlapped with the electrical energy. The overlapping factors of two cross sections are calculated to be 6.28% and 24.3%, respectively. It is very clear that the subwavelength silicon bridge (72 nm wide) between air holes of the PC nanocavity helps to improve the overlapping factor.

For resonator-based E-O modulators, the modulation speed depends on two factors: the photon lifetime and the resistance-capacitance (RC) time delay constant. The overall 3dB bandwidth can be calculated as $f_{3dB} = 1/\left(1/f_{opt}^2 + 1/f_{RC}^2\right)^{1/2}$. The photon-lifetime-limited bandwidth can be calculated as $f_{opt} = f/Q$, where $f$ is the resonance frequency and Q is the Q factor of the resonator. For PC nanocavity, the Q factor can be controlled by the air holes in the PC mirror segments [30] and the doping level. In this design, we target a Q factor of 5,000, which won't limit the bandwidth up to 38 GHz at 1.55 μm wavelength. Therefore, the speed of the PC nanocavity modulator is majorly limited by the RC delay. However, a high capacitance density of the MOS capacitor is crucial to reduce the driving voltage and improve the energy efficiency [14]. Thus, it is especially important to reduce the series resistance in order to achieve both high-speed and high energy efficiency. The nanocavity modulator contains semiconductor conduction path through the silicon waveguide and thin ITO gate layer, which contribute to the majority of the series resistance. Thus, we optimized the electrical configuration, and reduced the resistance from both silicon and ITO. First, on the silicon side, the high-resistance silicon strip conduction path [10] is replaced by partially etched silicon slabs. Then, the doping of the silicon conduction path is carefully engineered. Fig 1d shows the partial layout used for the fabrication of the nanocavity modulator. The 50nm thick silicon slabs and the top 50nm-thick layer of the silicon waveguide (p+ region) are moderately doped, leaving the bottom 200nm-thick layer of the silicon waveguide lightly doped (p-). The top 50nm-thick layer of the silicon contact area (p++ region) is highly doped and extends to only 500nm away from the silicon waveguide edge. Fig 1c illustrates the cross-sectional doping profile in this design. Next, on the ITO side, the length of the TCO conduction path is reduced by patterning the metal contact pad 1 μm away to the nanocavity. In order to quantitatively evaluate the high-speed performance of the PC nanocavity modulator, we performed simulation of the modulator using ANSYS HFSS. The resistance and capacitance parameters are extracted according to the equivalent circuit model shown in Fig 1e. In the simulation, the dielectric constant of $HfO_2$ has been adjusted to match the 8 fF/μm$^2$ capacitance density of the ITO/16nm $HfO_2$/Si MOS capacitor simulated by quantum moment model [31]. We simulate both cases for the MOS capacitor covering the centric one and three periods of the PC nanocavity, which correspond to capacitances of 6.1 fF and 18.3 fF, respectively. The silicon series resistance can be divided into contributions from the three different doping regions (p-, p+, p++). Then, the total series resistance R can be expressed as:

$$\begin{aligned}R &= R_{Si} + R_{ITO} = (R_{Si,p-} + R_{Si,p+} + R_{Si,p++}) + R_{ITO} + R_r \\ &= R_{s,p-}A_{p-} + R_{s,p+}A_{p+} + R_{s,p++}A_{p++} + R_{s,ITO}A_{ITO} + R_r\end{aligned} \quad (3)$$

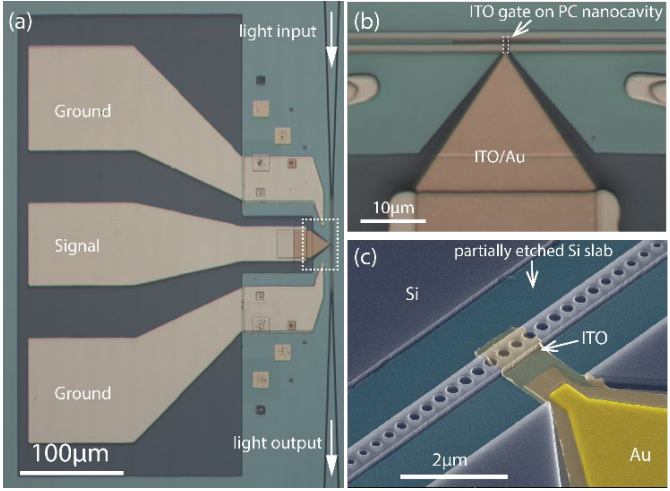

Fig. 4. (a) Optical image of the fabricated PC nanocavity modulator. (b) Zoomed-in optical image of the nanocavity modulator at the active region (dashed box in (a)). (c) Colored scanning electron micrograph (SEM) of the active region of the modulator.

Here $R_s$ and A are the sheet resistance and the effective aspect ratio of different regions, and $R_r$ is the remaining resistance that can't fit into the simplified linear resistance model. The effective aspect ratio is fitted by changing the sheet resistance of different regions. Detailed extracted parameters are listed in Table I. In our design, we choose doping concentrations of $1×10^{17}$ cm$^{-3}$, $5×10^{18}$ cm$^{-3}$, $1×10^{20}$ cm$^{-3}$ for p-, p+ and p++ regions of silicon, and $2×10^{20}$ cm$^{-3}$ carrier density for ITO. The corresponding S11 parameters from the HFSS simulation is plotted in Fig 3a. It gives us a total series resistance of ~3720 Ω and ~2250 Ω for 1 period and 3 periods coverage of the MOS capacitor, yielding RC-limited bandwidth of 7GHz and 3.9GHz, respectively. Fig 3b shows the normalized voltage across the MOS capacitor $V_{MOS}$ as a function of frequency by assuming 1V at the signal input. It equals to 2V at low frequency due to the signal reflection. We have verified the electrical configuration with FDTD simulation that with such doping we can still achieve a Q factor of ~5,000.

The nanocavity modulator is fabricated on a 250 nm thick lightly doped p-type silicon-on-insulator (SOI) substrate. First, the waveguide and PC nanocavity are patterned by electron-beam lithography (EBL), followed by reactive ion etching (RIE) of 200 nm thickness into the silicon, leaving a 50 nm thick Si slab for conduction and PC nanocavity with partially etched air holes. The grating couplers are pattern by a second round of EBL and RIE. Then, the p+ region and p++ region is selectively implanted with 5 keV B+ ions at flux of $2.5×10^{13}$ cm$^{-2}$ and $6×10^{14}$ cm$^{-2}$ with tilted angles of 7° and 30°, respectively. After the ion implantation, the dopants are activated by rapid thermal annealing at 1000 °C for 10 seconds. Next, 16nm thick $HfO_2$ layer is deposited using atomic layer deposition (ALD). After the ALD, the oxide at the silicon contact region is removed by buffered hydrofluoric acid. Ni/Au coplanar ground-signal-ground (GSG) electrode pads are patterned by photolithography, thermal evaporation and lift-off. Finally, the Au contact pads under the ITO and the ITO gate layer are patterned by EBL and lift-off. 20 nm of ITO gate layer is sputtered at 1% $O_2$/Ar gas flow, which yields an ITO resistivity of ~$1.4×10^{-3}$ Ω·cm. The fabricated device is shown in Fig 4a and 4b. The length of the ITO gate is 1 μm, which covers the centric three periods of the PC nanocavity as shown in Fig 4c.

### III. EXPERIMENT RESULTS

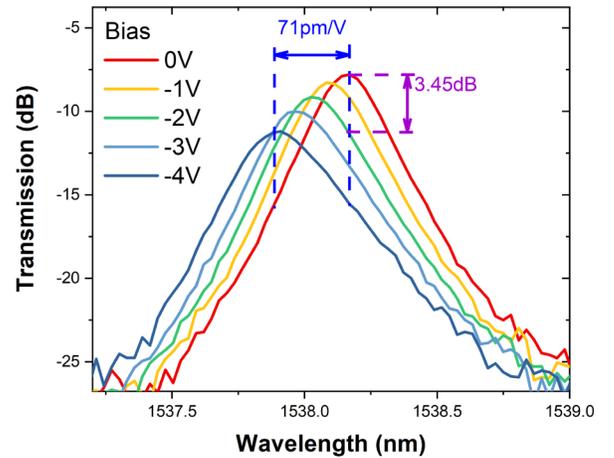

Fig. 5. Transmission spectra of one fabricated PC nanocavity modulator at different bias voltage.

The fabricated devices are characterized with light input and output via surface grating couplers. A polarization controller is used to excite the transverse electric (TE) mode light. The PC nanocavities are designed to operate at C-band. Fig 5 shows the transmission spectra at different bias voltages of a fabricated nanocavity modulator with resonance wavelength at 1538 nm.

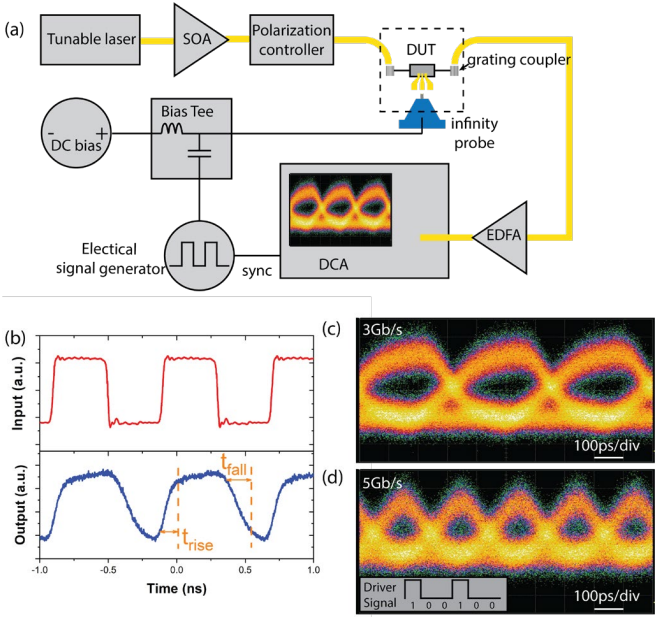

Fig. 6. (a) Schematic of the AC testing setup. (b) Electrical input and optical output AC modulation signal at 1.25 GHz with 2 $V_{pp}$ voltage swing. (c) and (d) Measured signals at 3 Gb/s and 5 Gb/s, respectively.

The plotted transmission is normalized to that of a straight waveguide with the same grating couplers. The device has 22 air holes in each PC mirror segments, which corresponds to a total device length of 14.8 μm. Considering the partially etched silicon slab width of 1.2 μm, the device occupies a compact footprint of only 43 μm². The PC nanocavity exhibits a Q factor of 9,460 before doping and ITO gate deposition. After the device fabrication, the Q factor is reduced to ~5,600, which is majorly due to the silicon doping. Insertion loss (IL) of the nanocavity is 7.8 dB, which is primarily limited by the fabrication imperfection of EBL and RIE process. With better control of fabrication processes, the IL can potentially be reduced to less than 0.5 dB [8]. With -4 V applied bias, the peak resonance wavelength shifts 284 pm, corresponding to a wavelength tunability of 71 pm/V. At -2 V bias, a peak extinction ratio (ER) of 3.45 dB can be reached. The wavelength tunability is less than the previous demonstration [10] due to two reasons. First, the HfO$_2$ gate oxide layer is thicker than the previous case, which results in smaller capacitance density of the MOS capacitor. Second, limited by the alignment accuracy of our EBL instrument, we have to pattern and etch the air holes and silicon slabs in the same step of fabrication process. This leads to under etching of air holes due to the local loading effect of the RIE [32]. As a result, the real overlapping factor is smaller than the previous ideal estimation, since more optical field is contained inside the silicon waveguide slab. The overlapping factor can be esimated as

$$\alpha = \frac{\Delta\lambda}{V} \frac{V_m}{\lambda \cdot K \cdot C}, \qquad (4)$$

where Δλ/V is the wavelength tunability, λ is the resonance wavelength, and C is the capacitance. The average free carrier dispersion coefficient can be estimated, as $(K_{p-Si}+K_{ITO})/2$, to be ~8.1×10$^{-3}$ C·cm³ [14]. Then, we can calculate the actual overlapping factor to be 1.79%. As a comparison, our previous demonstration achieved a wavelength tunability of 250 pm/V with 1 period coverage of the MOS capacitor and an overlapping factor of 7.43%, which is very close to the overlapping estimation in the previous section. Therefore, optimizing the fabrication processes to achieve a similar overlapping factor for high speed E-O modulators is still necessary for future research.

The high speed response of the nanocavity modulator is measured using the setup shown in Fig 6a. The AC signal is generated by Anritsu MP1763B and combined with a DC bias through a bias tee before applying to the device through a high-speed probe (Infinity Probe). The output optical signal is measured by a digital communications analyzer (Agilent 86100A). Fig 6b plots the AC EO modulation at 1.25 GHz with 2 $V_{pp}$ rectangular voltage swing. The rising and falling time of the optical response are measured to be ~0.135 ns and ~0.18 ns, respectively, which corresponds to a 3-dB bandwidth of 2.2 GHz. The asymmetric waveform is due to the dynamics of resonator based modulator [33]. Assuming the capacitance of 18.3 fF from the previous simulation, we can calculate the series resistance of the nanocavity modulator to be ~3.9 kΩ. The ITO conduction path contributes ~0.6 kΩ to the series resistance, which can be calculated from the ITO sheet resistance. Then, ~3.3 kΩ comes from the silicon conduction path, which also matches the estimation from the resistance between the two ground electrode pads. Overall, the experiment result matches well with the HFSS simulation. Besides, the sheet resistance of the p++ region on the device chip is measured to be ~1100 Ω/□, which indicates that the larger resistance from the silicon conduction path may be due to inaccurate implantation conditions or partial activation of the dopants, and the induced large contact resistance. Finally, we measured the high-speed digital modulation of the nanocavity modulator up to 5 Gbps. Fig 6c and 6d show the measured signals at 3 Gbps and 5 Gbps data rate. The asymmetric shape of the signal is due to the input driven signal we used for the measurement, as is shown in the inset of Fig 6d.

IV. STRATEGIES FOR HIGHER MODULATION BANDWIDTH AND ATTO-JOULE PER BIT ENERGY EFFICIENCY

We have proved the capability of the gigahertz operation of the TCO-Si PC nanocavity modulator, which matches the prediction of our design and simulation. Here, we discuss how we can further improve the modulation speed. Basically, the series resistance from both the silicon and ITO need to be reduced. On the silicon side, the series resistance can be further reduced using node-matched doping technique [34]. Such method allows us to put the p++ doping region closer to the active cavity region without significantly affecting the optical mode as is illustrated in Fig 7 a. FDTD simulation also gives us a Q factor larger than 5,000. On the gate side, by replacing ITO ($\mu$=20 m$^2$/(Vs)) with high mobility TCO materials such as Ti-doped In$_2$O$_3$ ($\mu$>80 m$^2$/(Vs))[35], the TCO series resistance can be reduced by more than 4×. In the meantime, larger mobility also induces less plasma absorption. We simulated the resistance of the proposed structure. It yields a total series resistance of less than 852 Ω, which leads to a RC bandwidth of 30 GHz for 1 period coverage of MOS capacitor. The overall 3dB bandwidth reaches over 23.5 GHz. Such optimized nanocavity modulator can be modulated at 40 Gbps using simple OOK for on-chip optical interconnects.

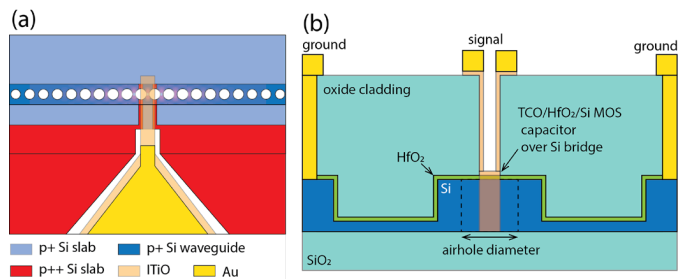

Fig. 7 (a) Layout schematic of node-matched doping profile with overlay of the electrical energy density distribution. (b) Cross-sectional schematic for building TCO-Si MOS capacitor over the subwavelength Si bridge through making window on the top cladding layer.

The energy consumption of this PC nanocavity modulator can be calculated as CV$^2$/4, to be 18.3 fJ/bit by assuming 2 V$_{pp}$. However, we anticipate that larger wavelength tunability, smaller driving voltage, and higher energy efficiency can be achieved. For example, 285 pm/V wavelength tunability should be achievable by increasing the overlapping factor from 1.79% to 7.2%. Based on the Q factor and capacitance of current device, the driving voltage can be reduced to 0.5 V, leading to energy consumption of 1.14 fJ/bit. Furthermore, as discussed in Section II, the subwavelength silicon bridge between air holes exhibits much larger overlapping factor than the overall PC nanocavity. By principle, we can build the active TCO-Si MOS capacitor only at the subwavelength bridge region, for example, through making window on top of the cladding layer as shown in Fig. 7b. Assuming we can increase the overlapping factor from 7.2% to ~22%, which corresponds to 100 nm long region of the silicon subwavelength bridges (green dashed box in inset of Fig 2d), we can further reduce the required capacitance by nearly 3× to 6 fF, while not sacrificing the wavelength tunability. In that case, the energy consumption can potentially be reduced to 375 aJ/bit.

Finally, we briefly compare the energy efficiency of our TCO-Si PC nanocavity modulator with other silicon resonator-based E-O modulators based on different resonator designs and capacitors. The energy consumption of a resonator-based modulator is dependent on four main factors [14]: the carrier dispersion coefficient K of the active material, the Purcell factor of the resonator F$_P$, the total capacitance C, and the overlapping factor α. The comparison of these devices is listed in Table II. We can see that there are two main advantages of the TCO-Si PC nanocavity modulator. First, the small mode volume of PC nanocavity helps to achieve large Purcell factors. For high-speed resonator-based modulators, the Q factor is typically limited to only a few thousand, due to photon-lifetime bandwidth consideration. Then, a small mode volume becomes especially important. In this work, we achieve a Purcell factor of PC nanocavity almost an order of magnitude larger than that of mico-disk modulators. Second, the large capacitance density and design freedom of the MOS capacitor help us to achieve a reasonable capacitance in a miniaturized active device volume. The main drawback of the TCO-Si hybrid PC nanocavity modulator is the relatively small overlapping factor. The carrier accumulation of MOS capacitors occurs at the interfaces of the waveguide, while the perturbation of free carriers of reversed biased PN junction is inside the waveguide, which intrinsically has better overlapping with the optical mode. For example, the overlapping factor of our previous PC nanocavity modulator is more than one order of magnitude smaller than that of a vertical PN junction micro-disk modulator [4], which reaches a near perfect overlapping of 94%. Luckily, the overlapping factor of the TCO-Si hybrid PC nanocavity modulator can be further optimized by taking advantage of the subwavelength structure of PC nanocavities as discussed here. Besides, further reducing the energy consumption can also possibly be achieved by adapting PC nanocavities with more miniaturized cavity mode volume, such as slot PC cavities [36] or bowtie PC cavities [37]. In summary, TCO-Si PC nanocavity modulator offers us the best possibility to achieve both high-speed operation and extremely low energy consumption at hundreds of atto-joule per bit level simultaneously. However, significant engineering optimization is still needed to achieve such ultimate goals.

TABLE II
COMPARISON OF SILICON RESONATOR-BASED MODULATORS

| Ref | Capacitor /resonator type | V$_m$ (($\lambda/n_{Si}$)$^3$) | gF$_p$[a] | C (fF) | α | E$_{bit}$ (fJ/bit) |
|---|---|---|---|---|---|---|
| [38] | vertical PN junction /micro-disk | 4.46 | 166 | 20 | 8% | 61 |
| [4] | vertical PN junction /micro-disk | 6.82 | 146.5 | 17 | 94% | 1 |
| [7] | lateral PN junction /micro-ring | 36.29 | 87.5 | 50 | 46% | 50 |
| [39] | Si/oxide/Si MOS /micro-ring | 21.20 | 28 | 320 | 23% | 180 |
| [40] | interleaver PN junction /micro-ring | 105.74 | 29 | 66 | 51% | 66 |
| [10] | hybrid Si-ITO MOS /PC nanocavity | 0.55 | 855 | 13 | 7.4% | 3.25 |
| proposal in this paper | hybrid Si-ITO MOS /PC nanocavity | 0.66 | 1247 | 6 | 22% | 0.375 |

[a] Purcell factor F$_P$ is defined as $F_P = (3/4\pi^2)(\lambda/n)^3 (Q/V_m)$. Detune factor g is defined as the ratio between the the full width at half maximum (FWHM) of the resonator to the required resonance detuning of the modulator.

## V. Conclusion

In summary, we presented comprehensive design and in-depth analysis of an ultra-compact, high-speed TCO-Si PC nanocavity modulator based on ITO/$HfO_2$/Si MOS capacitor. We quantitatively analyzed the overlapping factor between the accumulated free carriers and the cavity resonant mode. Its contribution to the energy efficiency of the modulator is sufficiently discussed. The relationship between the doping of semiconductor conduction path and modulation speed was systematically investigated via high frequency simulation. In our experimental demonstration, the fabricated TCO-Si nanocavity modulator shows a wavelength tunability of 71 pm/V with a Q factor of ~5,600, achieving 3.45 dB ER with 2 $V_{pp}$ applied bias. We demonstrated a 3dB modulation bandwidth of 2.2 GHz. E-O modulation was measured up to 5 Gb/s with 2 $V_{pp}$ voltage swing, which corresponds to an energy efficiency of ~18.3 fJ/bit. Besides, we proposed that the series resistance can be further reduced by node-matched doping of Si and high-mobility TCO material, which can improve the 3dB bandwidth to over 23.5 GHz. Furthermore, based on the quantitative analysis of the overlapping factor, we predict that single-digit femto-joule per bit energy efficiency of 1.14 fJ/bit can be reached through optimizing the gate oxide thickness and fabrication processes. Further increasing the overlapping factor is possible through precisely covering the TCO-Si MOS capacitor at the subwavelength silicon bridge of PC nanocavity. With that, an extremely low energy consumption of 375 aJ/bit can be achieved. As the first systematic analysis and experimental demonstration of high-speed Si-TCO resonator-based modulator, this work directly proves the potential of TCO-gated silicon photonic devices for high density, high-speed, ultra-low energy optical interconnect systems.

Acknowledgment

The authors would acknowledge the support from the EM Facility and MASC center at Oregon State University for the device fabrication.